\documentclass{emulateapj}
\usepackage{comment}
\usepackage{natbib}
\usepackage{xcolor}
\usepackage{amsmath}
\usepackage[backref,breaklinks,colorlinks,citecolor=blue]{hyperref}
\usepackage{times}

\newcommand{\sm}{\, {\rm M}_{\odot}}

\newcommand{\kms}{{\rm km/s}\,}

\def\gsim { \lower .75ex \hbox{$\sim$} \llap{\raise .27ex \hbox{$>$}} }
\def\lsim { \lower .75ex \hbox{$\sim$} \llap{\raise .27ex \hbox{$<$}} }

\newcommand{\cutt}[1]{}

\shorttitle{Evolution of gaps in streams} 
\shortauthors{A. Helmi and H.H. Koppelman}

\begin{document}


\title{The time evolution of gaps in tidal streams}

\author{Amina Helmi and Helmer H. Koppelman}
\affil{Kapteyn Astronomical Institute, University of Groningen, P.O. Box 800, 9700 AV Groningen, The Netherlands}
\email{e-mail:ahelmi@astro.rug.nl}

\begin{abstract}
  We model the time evolution of gaps in tidal streams caused by the
  impact of a dark matter subhalo, while both orbit a spherical
  gravitational potential. To this end, we make use of the simple
  behaviour of orbits in action-angle space. A gap effectively results
  from the divergence of two nearby orbits whose initial phase-space
  separation is, for very cold thin streams, largely given by the
  impulse induced by the subhalo. We find that in a spherical
  potential the size of a gap increases linearly with time for
  sufficiently long timescales. We have derived an analytic expression
  that shows how the growth rate depends on the mass of the perturbing
  subhalo, its scale and its relative velocity with respect to the
  stream. We have verified these scalings using N-body simulations and
  find excellent agreement. For example, a subhalo of mass 10$^8 \sm$
  directly impacting a very cold thin stream on an inclined orbit can
  induce a gap that may reach a size of several tens of kpc after a
  few Gyr. The gap size fluctuates importantly with phase on the
  orbit, and it is largest close to pericentre. This indicates that it
  may not be fully straightforward to invert the spectrum of gaps
  present in a stream to recover the mass spectrum of the subhalos.
 \end{abstract}


\section{Introduction}
\label{sec:intro}

A key prediction of the concordance cold dark matter model of
structure formation is the presence of myriads of dark satellites
orbiting the halos of galaxies like the Milky Way \citep{klypin1999,moore1999}.  The presence of
these objects is directly related to the fundamental nature of the
dark matter particle, and hence it is of uttermost importance to
establish if such subhalos indeed they exist, as well as their abundance and
properties.

Because such subhalos must be devoid of stars, they are very difficult to detect
and the only way in fact, may be through their gravitational influence.
Gravitational lensing is one of the means to detect their presence, although
this technique may be only realistically sensitive to the largest
subhalos \citep{Vegetti2014}. A powerful alternative is to measure their impact on stellar
streams orbiting the halos of galaxies \citep{johnston2002,ibata2002}. Streams are composed effectively of
stars on very nearby orbits, and hence if a subhalo comes close to
such a stream, it will slightly modify the orbits of those stars
leading to a change in its structure and to the formation of a gap \citep{Yoon2011}. 

It has been argued that the distribution of gap sizes can be used to
infer the mass spectrum of perturbers, and this is a truly interesting
prospect \citep{Carlberg2009,CG2013,EB2015b,Bovy2016}. Most works so
far have explored circular orbits for the streams as they move in a
spherical potential \citep{Carlberg2013,EB2015,EB2015b,E2016},
although \citet{Carlberg2015} has considered the
effect of eccentricity on gaps in streams orbiting in a triaxial mass distribution. Most recently, \citet{Sanders2016}
have modeled the evolution of a gap in a stream on a non-circular orbit
in an axisymmetric potential, but their focus has been on the
behaviour in angle and frequency space. At the moment no simple
analytic model exists that can predict how a gap once formed, it
evolves with time in {\it physical} space, and how its characteristics
depend exactly on the properties of the subhalo and the
encounter. This is in fact the goal of this Letter. It may be seen as
an important step to a full modelling of the gaps spectrum in a
cosmological context, for example along the lines of \citet{E2016}.

The paper is organised as follows. In Sec.~\ref{sec:method} we describe the
method used to model the evolution of a stream and the N-body simulations
carried out to validate the approach. In Sec ~\ref{sec:results} we describe the results
and we conclude in Sec ~\ref{sec:concl}.

\section{Methods}
\label{sec:method}

The model we use is based on two ingredients: the use of the impulse
approximation, and the divergence of nearby orbits. We proceed to
describe these in what follows, and then the N-body simulations we
have used for validation.

\subsection{The impulse approximation}

The impulse approximation \citep[see Ch.8,][]{BT2008} can be used to determine the perturbation
induced by a subhalo on a stream star as well as its dependence on the
properties of the subhalo and the relative motion with respect to the stream.

We follow here the description by \citet{EB2015}, where the stream's
velocity at the position of impact with the subhalo is aligned with
the $y$-direction, implying this is also the direction of the stream
(locally). The stream moves in the $x-y$ plane with velocity $v_y$,
while the subhalo of mass $M_{s}$ has velocity $(w_x,w_y,w_z)$ at the
time of impact. For simplicity we assume that the subhalo crosses the
stream itself, i.e. the impact parameter is $b = 0$~kpc. Using the
impulse approximation, the change in each of the velocity components
for stars on the stream $v_i$ can be computed from
\[
\Delta v_i = \int_{-\infty}^{+\infty} a_i (x,y,z) {\rm d}t,
\]
where $a_i$ is the acceleration field in the $i$-direction due to the
subhalo on a star located at position $(x,y,z)$.  This expression
can be computed numerically for any functional form for the subhalo's
mass distribution \citep{Sanders2016} such as e.g. the cosmologically motivated (truncated)
NFW \citep{nfw}. However, for a Plummer sphere and assuming the stream is
1-dimensional (i.e. $x$ and $z$ are constant), it takes a particularly simple form:
\begin{eqnarray}
\Delta v_x &=& 2GM_{s} \frac{y w_\perp w_\parallel \sin \alpha}{w (r_s^2 w^2 + w_\perp^2 y^2)}, \\
\Delta v_y &=& - 2GM_{s} \frac{y w_\perp^2}{w (r_s^2 w^2 + w_\perp^2 y^2)}, \\
\Delta v_z &=& - 2GM_{s} \frac{y w_\perp w_\parallel \cos \alpha}{w (r_s^2 w^2 + w_\perp^2 y^2)}. 
\label{eq:kick}
\end{eqnarray}
Here $w_\parallel = v_y - w_y$, $w_x = - w_\perp \sin \alpha$, $w_z =
w_\perp \cos \alpha$ and therefore $w_\perp = (w_x^2 +
w_z^2)^{1/2}$. It is sometimes argued that $\Delta v_x$ and $\Delta v_z$ can be neglected \citep[see
e.g.][]{Yoon2011}, however, in what follows we consider the velocity
change in all directions \citep[as in e.g.][]{EB2015,Sanders2016}. These expressions show that the velocity kick received by a star depends on its distance from the point of impact, falling off as $1/y$ for sufficiently 
large distances, and reaching maximum amplitude at
\begin{equation}
y_{\rm max}  = \pm w/w_\perp r_s,
\label{eq:ymax}
\end{equation} 
with value 
\begin{equation}
\Delta v_y^{\rm max} = \mp GM_s w_\perp/(w^2 r_s).
\label{eq:vymax}
\end{equation}
For a non-zero impact parameter $b$ the last expressions remain similar, with $r_s$ replaced by $(r_s^2+b^2)^{1/2}$. 

\subsection{The divergence of nearby orbits}
\label{sec:div}

Once a subhalo has given an impulse to stars located in a portion of a
stream, these will continue to orbit the host gravitational
potential, albeit on slightly modified trajectories. These
trajectories will diverge from each other in a fashion
that can be described well using the action-angle formalism \citep{hw99,HG2007}. 

We can use the behaviour of the spatial separation $\Delta {\bf X}$ of two nearby
orbits to actually describe the evolution of a
gap. On each side of the gap, we can imagine there being two stars $A$ and $B$ moving 
on different orbits that are slightly offset in position and velocity largely 
because of the kick received by the encounter with the subhalo. If their initial separation is $\Delta {\bf X}_0$ and $\Delta {\bf V} _0$, this
can be expressed in action-angle coordinates as 
\begin{equation}
\label{eq:delta0}
\left[ \begin{array}{c} \Delta {\bf \Theta}_0 \\
\Delta {\bf  J}_0 \end{array} \right]  
= M_0 \left[ \begin{array}{c} \Delta {\bf  X}_0 \\  \Delta {\bf  V}_0 \end{array} \right],
\end{equation}
where $M_0$ is the Jacobian matrix of the transformation from physical
and velocity space to action-angle space, i.e. $M_0 = \partial({\bf
  \Theta}, {\bf J})/\partial({\bf X},{\bf V})$ evaluated at the time
of the encounter, or minimum impact parameter, $t_0$, at the
phase-space location of e.g. star $A$. Recall that
\[\Delta {\bf  \Theta} = \Delta {\bf  \Theta}_0 + {\bf \Omega(J)} t, \qquad {\rm and} \qquad {\bf J} = {\bf J}_0,\]
where ${\bf \Omega(J)}$ are the frequencies of motion of e.g. star $A$'s orbit, or in matrix form
\begin{equation}
\label{eq:evol}
\left[ \begin{array}{c} 
\Delta {\bf  \Theta} \\
\Delta {\bf  J}  \end{array} \right]  
= \Omega' \left[ \begin{array}{c} \Delta {\bf  \Theta}_0 \\\Delta {\bf  J}  \end{array} \right],  
\end{equation}
with
\begin{equation}
\label{eq:omega_p}
\Omega' = \left[ \begin{array}{cc} I_3 & \partial {\bf \Omega} /\partial {\bf J}\, t\\
    0 & I_3 \end{array} \right],  
\end{equation}
where $\partial {\bf \Omega} /\partial {\bf J}$ is a 3$\times$3 matrix, also equal to the Hessian of the Hamiltonian. Furthermore, performing a local transformation, we find that at time $t$
\begin{equation}
\label{eq:delta}
\left[ \begin{array}{c} \Delta {\bf  X} \\  \Delta {\bf  V} \end{array} \right] = M^{-1}_t\left[ \begin{array}{c} 
\Delta {\bf  \Theta} \\
\Delta {\bf  J}  \end{array} \right]. 
\end{equation}
\\
Combining  Eqs.~(\ref{eq:delta0}), (\ref{eq:evol}) and (\ref{eq:delta}) we finally obtain 
\begin{equation}
\label{eq:deltat}
\left[ \begin{array}{c} \Delta {\bf  X} \\  \Delta {\bf  V} \end{array} \right] = M^{-1}_t \Omega' M_0 \left[ \begin{array}{c} \Delta {\bf  X}_0 \\  \Delta {\bf  V }_0 \end{array} \right],
\end{equation}
which allows us to measure the physical separation $\Delta {\bf  X}$ at time $t$ between nearby orbits, or in our case, the size of the gap
at any point in time. 

Let us consider what this predicts for sufficiently long timescales,
and for evolution in a spherical potential. In that case, the
motion occurs in a plane. This simplifies somewhat the matrices, in
the sense that they are either $2\times2$ or $4\times4$. The
spatial separation is given by $|\Delta {\bf X}|$ or $(\Delta {\bf
  X}^\dagger \Delta {\bf X})^{1/2}$, and this can be computed noting
that in Eq.~(\ref{eq:omega_p}), the dominant submatrix is the upper right one: $[\partial {\bf \Omega} /\partial {\bf J}\, t]$. Therefore in Eq.~(\ref{eq:deltat}): 
\begin{equation}
\Delta {\bf  X} \sim t \, M^{-1}_{t,1} [\partial {\bf \Omega} /\partial {\bf J}] \Delta {\bf  J}_0,
\label{eq:delta_x}
\end{equation}
where $M_{t,1}^{-1}$ is the upper left submatrix of $M_t^{-1}$, and transforms from physical to angle coordinates: $\partial {\bf X}/\partial {\bf \Theta}$. Therefore
\begin{equation}
|\Delta {\bf  X}| \sim t (\Delta {\bf  J}_0^\dagger C_{\bf x,\Omega} \Delta {\bf  J}_0)^{1/2},
\end{equation}
where $C_{\bf x,\Omega} = [\partial {\bf \Omega} /\partial {\bf J}]
(M^{-1}_{t,1})^\dagger M^{-1}_{t,1}[\partial {\bf \Omega} /\partial
{\bf J}]$ is a symmetric matrix that thus depends on the location of
the gap along its orbit and the orbit itself through the frequency
derivatives.  This equation shows explicitly that the physical
separation between nearby orbits, or equivalently, the size of a gap
increases linearly with time for long timescales. Since the matrix
$M^{-1}_{t,1}$ depends on the location of the gap at time $t$, this
shows also that the physical size of a gap will vary depending on its
orbital phase. From Eq.~(\ref{eq:delta_x}) we can also estimate the gap's volume as
$\propto \Delta X \Delta Y \propto t^2$ for a non-circular orbit in a spherical potential.  
More generally, the gap's volume will grow as $t^n$, with $n$ the number of independent frequencies (as encoded in the matrix 
$[\partial {\bf \Omega} /\partial {\bf J}]$).

Let us explore now the dependence of gap size on the subhalo mass and size and conditions of the encounter, all of which
are implicit in $\Delta {\bf J}_0$. The initial action
separation depends on $\Delta {\bf X}_0$ and $\Delta {\bf V}_0$, but
the term that dominates is that associated with the change in velocity
(Eq.~\ref{eq:vymax}). With the geometry considered for the
encounter, it can be shown using Eq.~(\ref{eq:delta0})  that $\Delta {\bf J}_0 = 2 \Delta v_y^{\rm max} 
{\bf f}_{\rm orb,0}$, where ${\bf f}_{\rm orb,0} = [x_0,  (v_{y,0} -
x_0\Omega_\phi)/\Omega_r]$, with $v_{y,0}$ the velocity of the stream
at the time and location of the impact and $x_0$ its $x$-location. The frequencies $\Omega_r$ and $\Omega_\phi$ are the radial and azimuthal frequencies of e.g. star $A$. 
Combining these expressions we find that for sufficiently long timescales, the gap size grows as 
\begin{equation}
|\Delta {\bf  X}| \sim t \frac{2 GM_s w_\perp}{w^2 (r_s^2 + b^2)}  (f^\dagger_{\rm orb,0} C_{\bf x,\Omega} f_{\rm orb,0})^{1/2}, 
\end{equation}
while the general expression for the gap's size at any point in time can be exactly determined using Eq.~(\ref{eq:deltat}). 

\begin{figure}
\includegraphics[width=9.5cm]{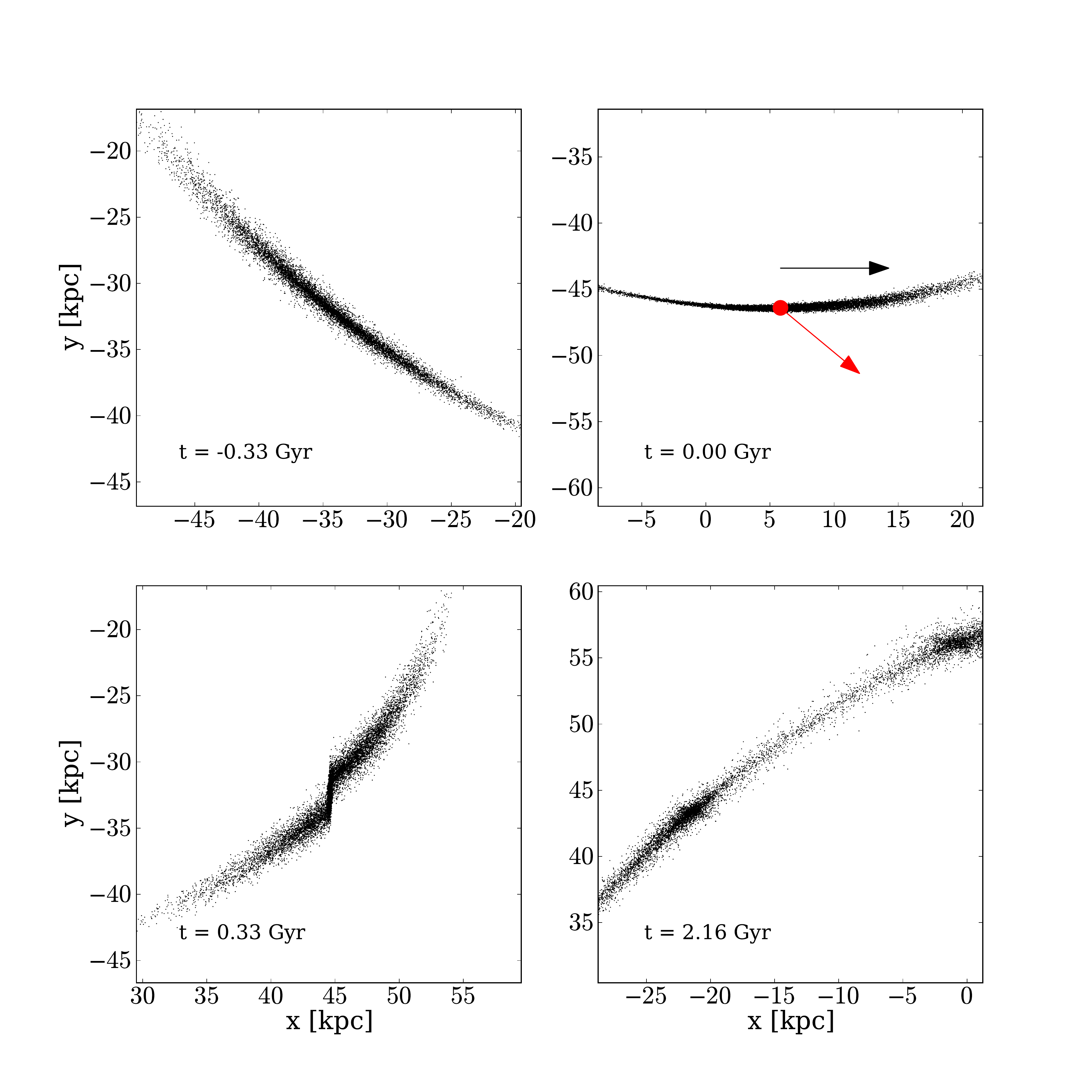}
\caption{Stream particles, before, during and after a direct encounter with a dark subhalo of $M_s = 10^{7.5} \sm$ (red symbol in the top right panel). At late times, a gap, in the form of a significant decrease in density, along the stream is clearly apparent. The arrows indicate the direction of the velocity vectors of the stream (black) and the subhalo (red) at the time of the encounter.}
\label{fig:nbody}
\end{figure}

\subsection{N-body simulations}

To validate the above analytic description, we have performed N-body
simulations of the encounters of a subhalo with a stream orbiting a
spherical NFW potential, of virial mass $M_{\rm halo} = 3 \times 10^{11} \sm$ and scale radius $r_s = 15.6$~kpc.

The progenitor of the stream is initially distributed following a
Gaussian in configuration and velocity space, with 1D-dispersions
$\sigma_x = 0.05$~kpc and $\sigma_v = 2$~kpc/Gyr ($\sim 2.04$ km/s),
respectively. It is evolved using {\sc Gadget-2} and
placed on an eccentric orbit with pericentre $r_p \sim 46$~kpc and
apocentre $r_a \sim 71$~kpc, for a total of $\sim 9$ Gyr.

At time $t=2.33$~Gyr the stream experiences an encounter with
a subhalo. This is modeled as a rigid Plummer sphere, i.e. we do not use
particles to follow its evolution. We have carried out experiments
using a range of masses and scale radii $(\log_{10}M_s[\sm], r_s [{\rm kpc}]) = [(6.9, 0.38), (7.2, 0.59), (7.5, 0.9), (7.9, 1.35)]$. All encounters have the same impact parameter $b\!=\!0$~kpc and the subhalo moves
with velocity $(w_x,w_y,w_z) = (80.1,  97.3, -23)~\kms$ in the frame in which 
the stream is on the $x-y$ plane, and the $y$-direction is aligned with the stream at the
time of impact, i.e. this the configuration used to computed the kicks
in Eqs.~(\ref{eq:kick}). At the time and location of the impact, the stream's velocity is $137.1~\kms$.

Figure \ref{fig:nbody} shows the stream before, during
and after an encounter with a subhalo of mass $M_s = 10^{7.5} \sm$ and
$r_s = 0.9$ kpc. The perturbation induced by the subhalo is clearly
apparent, and leads to the formation of a gap easily distinguished and
extending by more than $\sim 15$~kpc only 2~Gyr after 
the encounter.

\section{Results}
\label{sec:results}

\begin{figure}
\includegraphics[width=9.5cm]{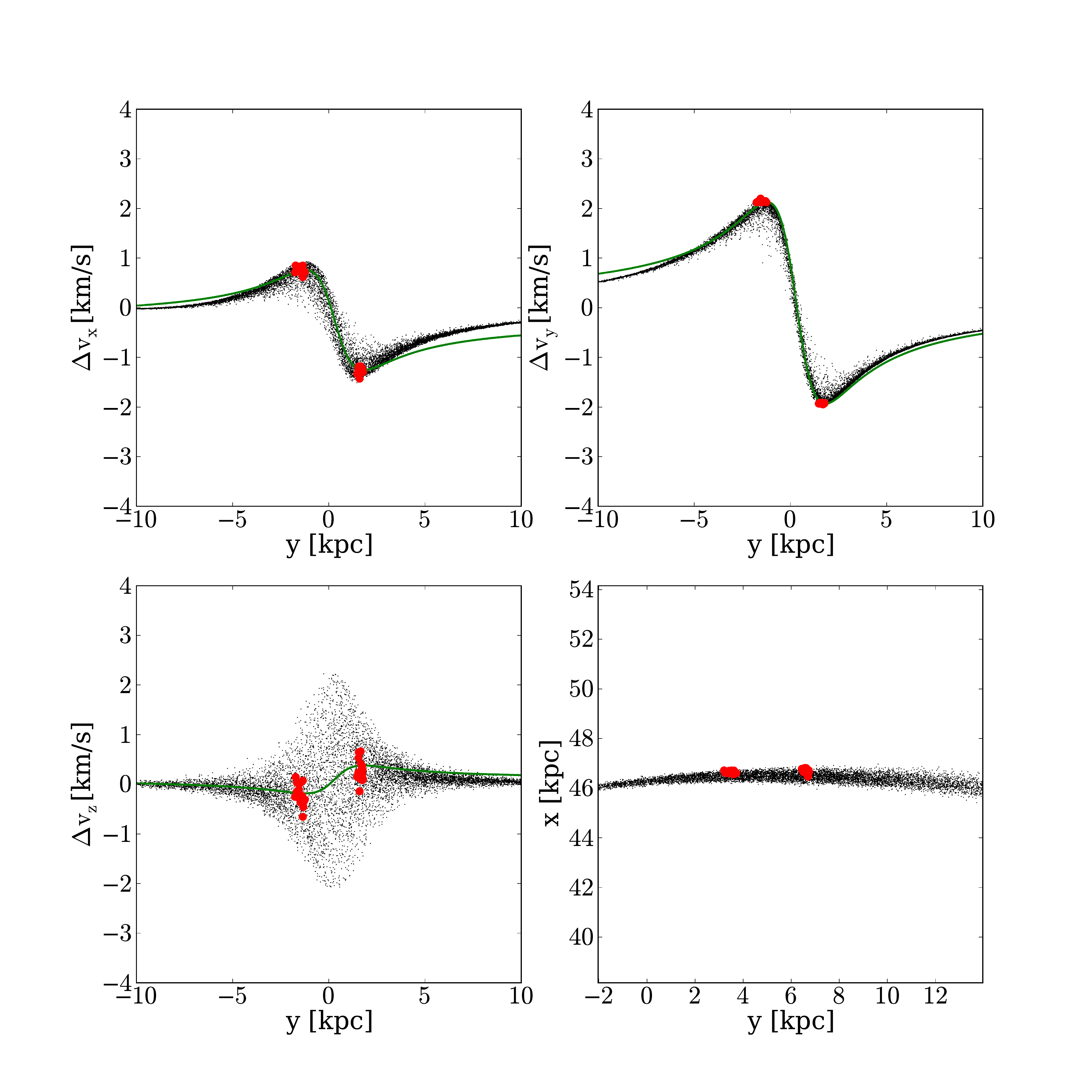}
\caption{Velocity changes along the stream for the experiment shown in
  Fig.~\ref{fig:nbody}. The
  red particles have been identified as those that have experienced
  the maximum velocity change.  The green curve
  indicates the prediction using the impulse
  approximation. \label{fig:impact}}
\end{figure} 

Fig.~\ref{fig:impact} shows the velocity change experienced by the
stream particles at the time of the collision for the experiment 
in Fig.~\ref{fig:nbody}. The solid curve corresponds to the
predictions from the impulse approximation, i.e.\
Eqs.~(\ref{eq:kick}), and they reproduce well the amplitude
and location of the maximum kick received by the stream
particles. The deviations at large distances can be attributed to the
stream's curvature \citep[see][]{Sanders2016}.  The coloured points
denote ``trailing'' and ``leading'' particles, i.e.\ located on either
side of the point of impact and that have experienced the maximum
velocity change, and which with time, will be on either side of the gap
that grows as a result of the encounter.

\begin{figure}
\includegraphics[width=9.5cm]{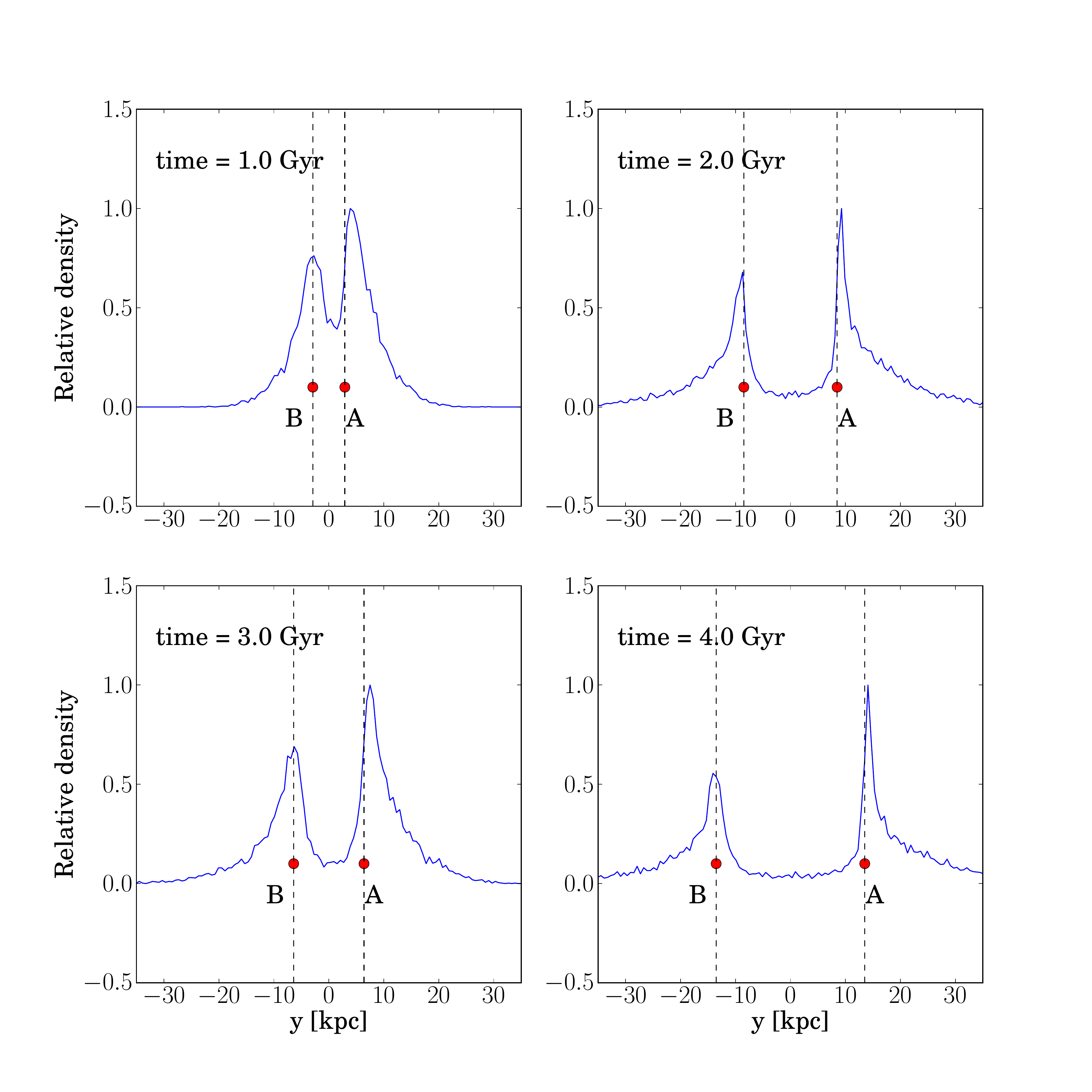}
\caption{Relative density along the stream around the location of the gap formed in the experiment in
  Fig.~\ref{fig:nbody}. The particles $A$ and $B$ are those that have experienced
  the maximum velocity change and are used to
  determine how the gap size evolves with time. \label{fig:size}}
\end{figure} 

This is explicitly shown in Fig.~\ref{fig:size} which depicts the
  density of the stream in the gaps' vicinity. The vertical lines in
  this figure indicate the location of the ``trailing'' and
  ``leading'' particles, and show that their separation follows well
  that of the density peaks around the gap at all times. We therefore,
  for computational ease, measure the gap size using the physical
  distance between these particles. Possibly such a position (and
  velocity) difference between two (groups of) star particles could be
  measurable with Gaia and follow-up spectroscopy, allowing direct
  comparisons to models.  Note that our method to measure the gap's
  extent differs from that of \citet{EB2015} who use the size of the
  underdense region. The two methods yield comparable physical extents
  when applied to our N-body simulations, with the gap size defined by
  the separation of the particles being only slightly larger, as can
  be seen from Fig.~\ref{fig:size}. 

\begin{figure}
\includegraphics[width=9cm]{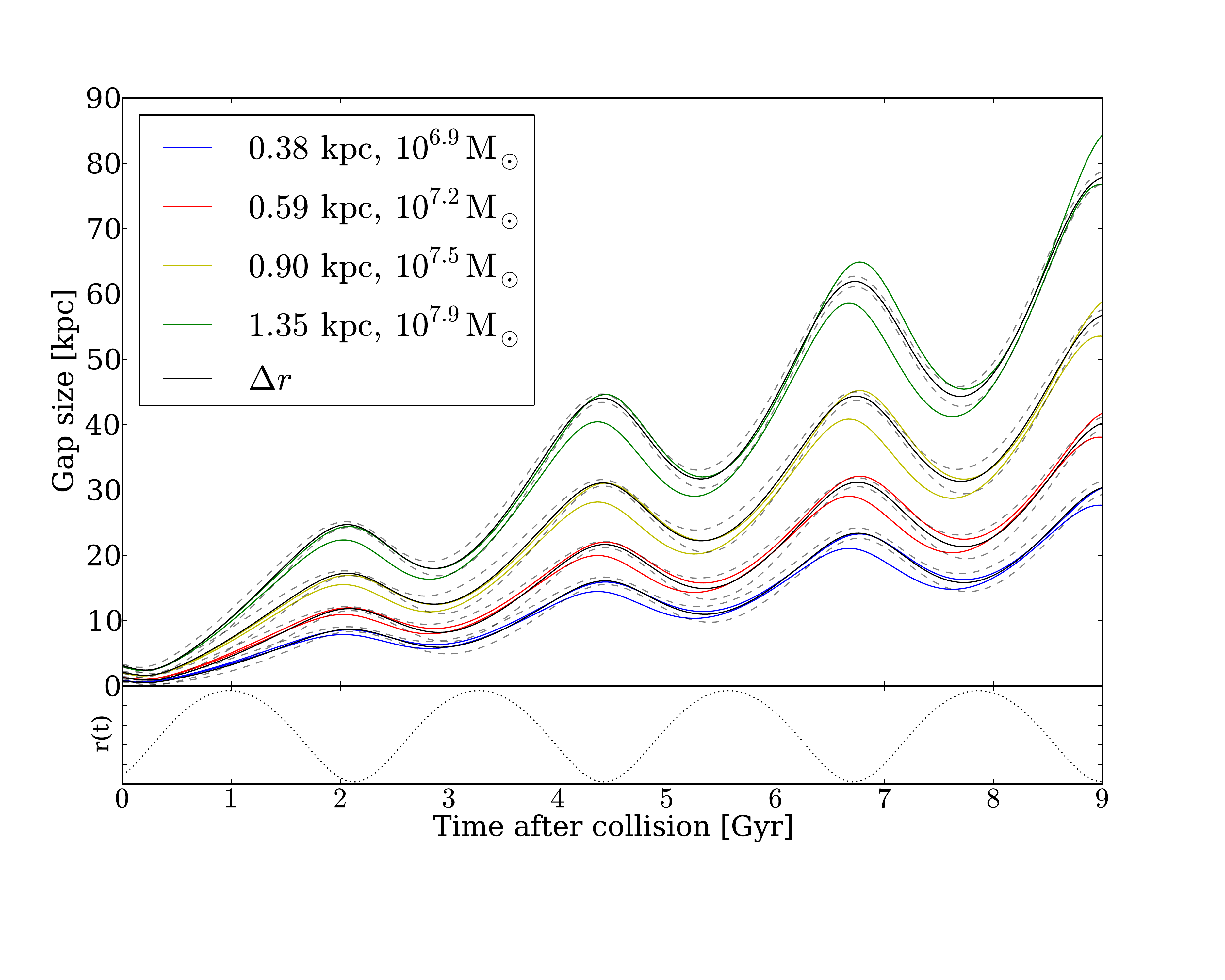}
\caption{The top panel shows the evolution of the gap size as function of time for encounters
  with different subhalos, as indicated in the inset. The
  agreement between the model predictions and the measured sizes is
  excellent. The bottom panel shows the radial orbital oscillation and evidences that the size
of a gap is largest when this is located close to pericentre. \label{fig:gaps}}
\end{figure}

Fig.~\ref{fig:gaps} shows the evolution of the gap size produced by
subhalos of different masses impacting the stream in the experiments
described above. For each experiment, the
average separation between pairs of ``trailing'' and ``leading''
particles is indicated with the black curve, while the dotted curves
correspond to the 1$\sigma$ scatter. The coloured curves in Fig.~\ref{fig:gaps} are the predictions
obtained using the formalism described in Sec.~\ref{sec:div}. Each
pair of coloured curves correspond to the separations $|\Delta {\bf
  X}|$ computed through linear perturbations around the orbits of
particles initially located on each side of the point of impact
(i.e. ``stars'' $A$ and $B$ of Sec.~\ref{sec:div}). The initial
separation $\Delta {\bf X}_0$ we take to be arbitrarily small and in
practise we set $\Delta x_0 = \Delta z_0 = 2\times 10^{-5}$ kpc, while
$\Delta y_0 = 2 y_{\rm max}$ from Eq.~(\ref{eq:ymax}), in the reference
frame aligned with the stream. For the initial velocity separation
$\Delta {\bf V}_0$ we use the prediction from the model, as described
in Eqs.~(\ref{eq:kick}) at the maximum. To this impulse driven
velocity change we add a term associated to the velocity gradient
${\bf \nabla_x V}$ along the stream over the volume $\Delta y_0$, which
is larger for larger subhalos (as $y_{\rm max}$ depends on $r_s$). The
velocity gradient is not exactly that given by the orbit of e.g. star
$A$ (as the stream does not follow a single orbit), but can be
computed using the formalism described in Sec.~\ref{sec:div} and in
particular using Eq.~(\ref{eq:deltat}) for arbitrary $\Delta {\bf
  X}_0$ and $\Delta {\bf V} _0$.  For the stream modeled, the velocity
difference due to the gradient is a factor 2 -- 4 smaller than the
impulse received along the direction of motion (but comparable or
larger in the other directions) as a consequence of the encounter with
the subhalos considered. This of course depends somewhat on the
specifics of the stream's progenitor orbit.

As shown in Fig.~\ref{fig:gaps} the agreement between the size of the
gap measured in the simulations and the predictions of our model is
excellent.  This implies that we are in a position
of predicting the size of a gap in a stream for any geometry, subhalo mass, scale
and density profile, at any point in time, for any stream orbiting a spherical potential.

As predicted by our model, the gap size oscillates strongly with time,
and comparison to the orbital radial oscillations plotted in the bottom
panel of the figure, shows that the gap is largest close to
pericentre.

\section{Discussion}
\label{sec:concl}

A gap in a stream is essentially the result of the divergence of
nearby orbits whose initial separation is driven by an encounter with
a dark matter subhalo. This conceptual framework allows us
to make detailed predictions for the evolution of gap sizes and their
dependence on the properties of the subhalos, the streams, their orbits
and the gravitational potential in which they move.

We have found that, for a spherical potential, gaps can grow very
fast, increasing their size linearly with time. Superposed on this
long-term behaviour, there are important oscillations that 
depend on their orbital phase. This long-term behaviour appears
to be in contrast to the $t^{0.5}$ growth proposed by \citet{EB2015} for upto 5 Gyr after
the encounter \citep[although][in their simulations also find linear
growth at late times]{Sanders2016}. Part of the difference, as
mentioned earlier, may lie in that we have considered general orbits
instead of only circular orbits. Additionally, differences in the orbital phase of the location of the encounter will lead
to different early-time behaviour. 

The important oscillations in gap size imply that one cannot infer the
mass of a subhalo directly from the size of a gap. For example,
Fig.~\ref{fig:gaps} shows that a gap of 10 kpc size could be induced
by a subhalo of mass $M_s \sim 10^{7.9} \sm$ less than 1 Gyr after
impact, but also by a subhalo with $M_s \sim 10^{6.9} \sm$ but 3.5 Gyr
after impact. This degeneracy comes on top of that identified by
\citet{EB2015b} between the mass of the subhalo $M_s$ and the impact
velocity $w$ (Eq.~\ref{eq:deltat}). Therefore, inferring the subhalo mass will strongly depend on
our ability to determine precisely the orbit of the stream in which
the gap is located. We have however, only focused on the spatial
characteristics of the gap, and not for example, on the kinematical
properties, which perhaps can help break some of the degeneracies
\citep[see][]{EB2015b}.  A statistical comparison of the predicted
  and observed distribution of gap sizes may also be a way to characterize the 
  granularity in the dark matter halos of galaxies \citep[see][]{Carlberg2012,E2016,Bovy2016}.

Although the gap size increases linearly with time, the volume it
  occupies will increase as $t^n$ with $n$ the number of independent
  frequencies of motion. For a general orbit in an arbitrary spherical
  potential $n=2$ while for a non-spherical potential there are at
  most 3 independent frequencies. This means that in this case, gaps
  may be more apparent since their internal density will be
  lower. The model we have developed is sufficiently general that it
can be applied in a statistical sense for an ensemble of
cosmologically motivated orbits and subhalo mass functions, an idea
recently put forward by \citet{E2016}. This will allow us to make
predictions specific to the $\Lambda$CDM model for the spectrum of
sizes of stream gaps for direct comparison to observations.

\acknowledgements

AH was partially supported by an NWO-VICI grant. We are grateful to
the referee for a constructive report, to Facundo G\'omez and Hans Buist for their contribution to improving
earlier versions of the software used for this Letter, and to Tjitske
Starkenburg with help in setting up the N-body simulations.

\end{document}